\documentclass[12pt,preprint]{aastex}

\newcommand{\hi}{H{\small ~I}}                   
\newcommand{\cv}{C{\small ~V}}                   
\newcommand{\cvi}{C{\small ~VI}}                 
\newcommand{\ovii}{O{\small ~VII}}               
\newcommand{\oviii}{O{\small ~VIII}}             
\newcommand{\fexvii}{Fe{\small ~XVII}}           
\newcommand{\neix}{Ne{\small ~IX}}               
\newcommand{\mgxi}{Mg{\small ~XI}}               
\newcommand{\oqkev}{$1\over4$~keV}
\newcommand{\tqkev}{$3\over4$~keV}

\shorttitle{X-ray Emission from the Heliosphere}
\shortauthors{Snowden, Collier & Kuntz}

\begin{document}

\title{{\it XMM-Newton} Observation of Solar Wind Charge Exchange Emission}

\author{S. L. Snowden\altaffilmark{1,2}}
\affil{Code 662, NASA/Goddard Space Flight Center, Greenbelt, MD 20771}

\author{M. R. Collier}
\affil{Code 692, NASA/Goddard Space Flight Center, Greenbelt, MD 20771}

\and

\author{K. D. Kuntz}
\affil{Astronomy Department, University of Maryland Baltimore County,
1000 Hilltop Circle, Baltimore, MD\ \ 21250}

\altaffiltext{1}{Universities Space Research Association}
\altaffiltext{2}{snowden@riva.gsfc.nasa.gov}
\begin{abstract}

We present an {\it XMM-Newton} spectrum of diffuse X-ray emission from 
within the solar system.  The spectrum is dominated by probable \cvi\ 
lines at 0.37~keV and 0.46~keV, an \ovii\ line at 0.56~keV, \oviii\ 
lines at 0.65~keV and $\sim0.8$~keV, \neix\ 
lines at $\sim0.92$~keV, and \mgxi\ lines at $\sim1.35$~keV.  This spectrum 
is consistent with that expected from charge exchange emission between 
the highly ionized solar wind and either interstellar neutrals in the 
heliosphere or material from Earth's exosphere.  The emission is clearly 
seen as a low-energy ($E<1.5$~keV) spectral enhancement in one of a series 
of four observations of the {\it Hubble} Deep Field North.  The X-ray 
enhancement is concurrent with an enhancement in the solar wind measured by 
{\it ACE}, {\it Wind}, and {\it SoHO} spacecraft.  The solar wind enhancement 
reaches a flux level an order of magnitude more intense than typical fluxes 
at 1~AU, and has a significantly enhanced O$^{+7}$/O$^{+6}$ ratio.  
Besides being of interest in its own right for studies of the solar 
system, this emission can have significant consequences for observations 
of cosmological objects.  It can provide emission lines at zero redshift 
which are of particular interest in studies of diffuse thermal emission 
(e.g., \ovii\ and \oviii), and which can therefore act as contamination in 
the spectra of objects which cover the entire detector field of view.  We 
propose the use of solar wind monitoring data as a diagnostic to screen 
for such possibilities.

\end{abstract}


\section{Introduction}

Diffuse X-ray emission from the solar system was clearly observed during 
the {\it ROSAT} All-Sky Survey (RASS) as a background component with 
temporal variations on scales from a large fraction of a day to many days.
These variations were dubbed Long-Term Enhancements (LTEs, \citet{sea95}), 
and provided a significant background particularly at \oqkev.  Although 
the origin of these LTEs at the time was unknown, the LTEs were associated 
with solar wind parameters \citep{frey94}.  With the observation of X-rays 
from comets \citep{lis96}, emission from solar wind charge exchange (SWCX) 
between solar wind ions and neutral material within 
the heliosphere was demonstrated \citep{cra97}.  SWCX emission  
subsequently was suggested as the origin for LTEs \citep{cox98,cra00}, and has 
recently been suggested as being responsible in quiescence for a significant 
fraction of diffuse X-ray emission at \oqkev\ \citep{lal04,rc03b} previously 
attributed to the Local Hot Bubble \citep{sea98}.  

More recently, evidence for geocoronal emission (SWCX with exospheric material)
has been detected in {\it Chandra} observations of the dark moon \citep{wea04}.  
This emission clearly originates in the near-Earth environment, which is 
consistent with the likely production of X-rays from the terrestrial 
magnetosheath \citep{rc03a}.  The statistics of the dark moon data are somewhat 
limited but they clearly show excess \ovii\ and \oviii\ emission.

Analysis of cometary X-ray spectra suggest that solar wind charge state
composition and speed affect the X-ray emission \citep{lea01,kea03,sc00}.  Thus 
through a combination of techniques, including X-ray and neutral atom imaging
among others, it may be possible to continuously monitor the solar wind from 
well inside Earth's magnetosphere.  However, the nature of solar system X-ray 
emission is strongly dependent on the point of view of the observer.  While 
the observation of SWCX X-ray emission samples aspects of the solar system of 
interest to astronomers, the emission can also provide a contaminating component 
which strongly impacts observations of extended cosmic X-ray sources.  The 
SWCX emission spectrum is dominated by highly ionized carbon, oxygen, 
neon, possibly iron, and magnesium lines, which are also of great 
astrophysical interest.  (The oxygen lines are particularly important as they 
are commonly used for temperature and density diagnostics of thermal emission 
from diffuse plasmas.)  
For objects at zero redshift (i.e., emission from the Milky Way 
and nearby galaxies) where the emission is expected to fill the entire field 
of view of the detector, SWCX emission is indistinguishable from that 
of the object, except to the extent that temporal variation can be detected.

While there is likely to be SWCX emission at some quiescent level
\citep{rc03b,lal04}, the strongest emission will be associated with flux 
enhancements of the solar wind.  These enhancements are long enough in duration 
that they can ``contaminate'' an entire observation, but variable enough that
if there are multiple observations of a single target they can be identified,
at least to some level.  This was the basis of the LTE ``cleaning'' of the 
RASS \citep{sea95,sea97}, and of {\it ROSAT} 
pointed observations which were typically distributed over the period of a 
few days if not weeks \citep{sea94}.

In this paper we report the detection of SWCX emission during an 
{\it XMM-Newton} observation of the {\it Hubble} Deep Field North.   We 
correlate that detection with a concurrent enhancement in the solar wind 
density observed by Advanced Composition Explorer ({\it ACE}), Solar and 
Heliospheric Observatory ({\it SoHO}), and {\it Wind}.  In \S~\ref{sec:data} 
we present the data, data reduction, and analysis, in \S~\ref{sec:discussion} 
we discuss the results, and in \S~\ref{sec:conclusions} we detail our 
conclusions both in regards to the emission itself and the implications for 
X-ray observations of more distant objects.

\section{Data and Data Analysis}
\label{sec:data}

\subsection{X-ray Data}

\subsubsection{Data Preparation}

The SWCX emission discussed in this paper was observed in the 
fourth (2001 June 1--2) {\it XMM-Newton} pointing of the {\it Hubble} Deep 
Field North (HDF-N, $\alpha=12^{hr}36^m50.00^s,\delta=+62^\circ13'00.0''$, 
a project Guaranteed Time observation).  Because of satellite constraints 
and the total length of the observation, it was broken into four segments 
(pointings), which were scheduled over a period of 16 days 
(Table~\ref{tbl:observations} gives the specifics of the pointings).
Because of this multiple coverage, the length of the individual pointings,
and relative freedom from other non-cosmic backgrounds (most critically 
the soft proton flaring), the HDF-N observation was chosen as a test case 
for a new non-cosmic background modeling and subtraction method for 
{\it XMM-Newton} EPIC data \citep{kea04}.  The serendipitous detection of 
the SWCX emission came from that testing.

\begin{figure}
\plotone{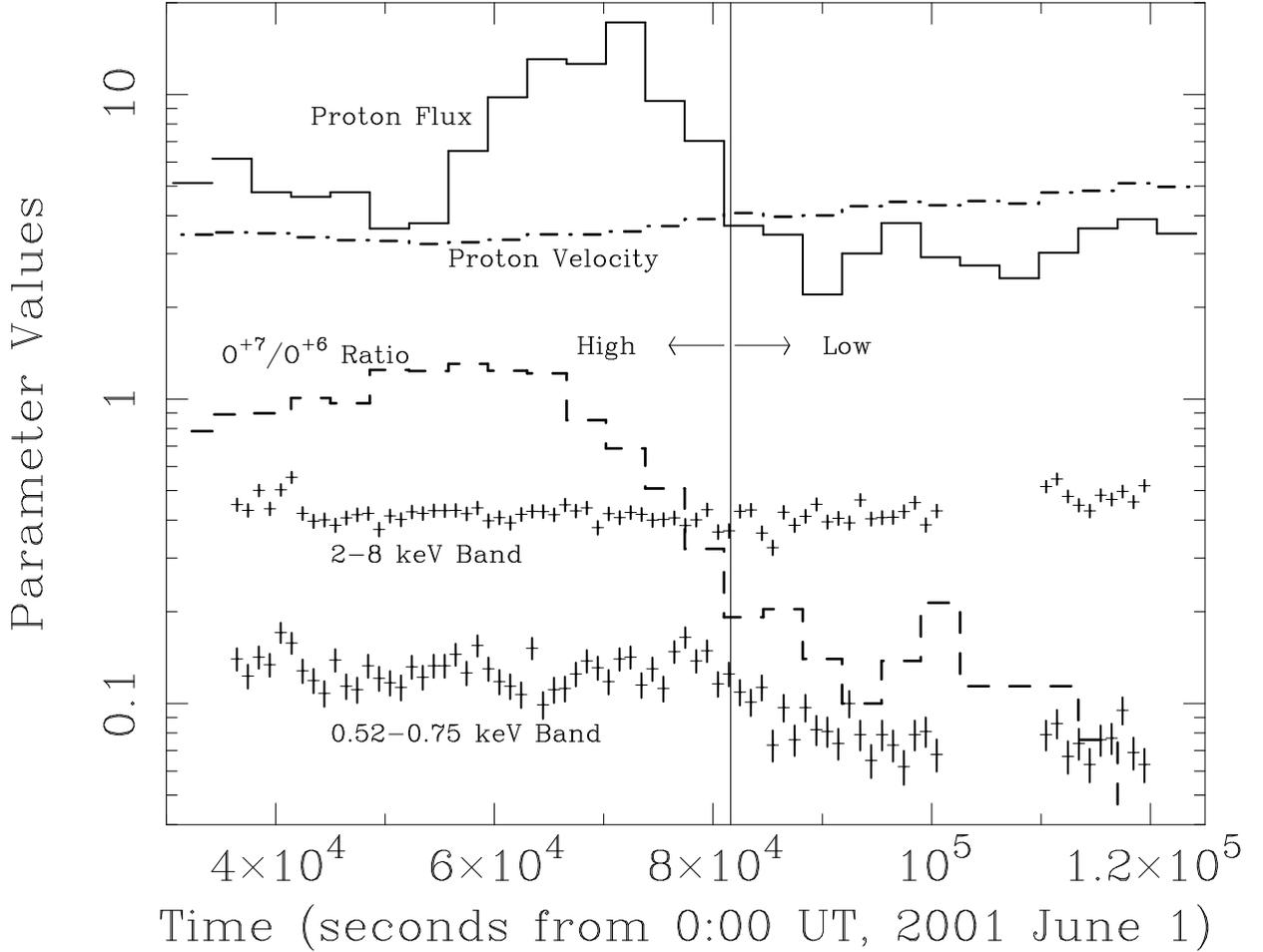}
\caption{Various parameter values for the observation affected by the 
SWCX emission (HDF-N \#4). The dot-dash curve shows the hourly 
{\it ACE SWEPAM} proton velocity data in units of 100~km~s$^{-1}$.  The 
solid curve shows the hourly {\it ACE SWEPAM} proton flux data in units 
of $10^8$~cm$^{-2}$~s$^{-1}$.  The dashed curve shows the hourly 
{\it ACE SWICS} O$^{+7}$/O$^{+6}$ ratio data smoothed with a five hour 
running average.  {\it ACE} data have not been adjusted for 
travel time from the L1 point to Earth (a delay of $\sim4000$~s).  
The upper set of points shows the {\it XMM-Newton} MOS1 
light curve for the 2--8~keV band in counts~s$^{-1}$.  The lower 
set of points shows the light curve for the 0.52-0.75 keV band 
in counts~s$^{-1}$, which includes the 
\ovii\ and most of the \oviii\ line emission. Periods 
of soft proton-flaring have been removed. The time used to separate  
the spectra with and without the SWCX emission is shown by the vertical
line.  The X-ray count rates do not have the particle background 
subtracted.
\label{fig:lc}}
\end{figure}

Outside of time intervals obviously affected by proton flaring (which
were removed for this analysis), the light curves for the observations 
were reasonably flat indicating at least a low if not zero level 
of residual flaring.  Figure~\ref{fig:lc} shows the X-ray light curves for 
the non-flaring periods of the fourth observation, and also includes 
the {\it ACE} data (discussed below).  The light curve for the 
2--8~keV band is relatively constant but the 0.52--0.75~keV band 
(covering the \ovii\ and most of the \oviii\ lines) shows
a significant drop in the last quarter of the interval.  We used
this light curve to separate the spectrum into high and low states at 
$T=81578$~s in Figure~\ref{fig:lc} (22:39:38 UT on 2001 June 1; 
$T_{XMM}=107822378$ seconds after 0~UT, 
1998 January 1, in the {\it XMM-Newton} time reference system).  
Table~\ref{tbl:observations} also lists the average count rates in the 
0.52--0.75~keV and 2--8~keV bands for the different pointings.  The
2--8~keV band count rate is reasonably consistent among all of the
pointings, as is the average count rate for the 0.52--0.75~keV band, 
except for the time period affected by the SWCX emission.  The 
minor, but greater-than-statistical scatter of the X-ray count rates 
in Table~\ref{tbl:observations} during the non-SWCX periods is likely 
due to the contribution of the unsubtracted particle 
background which is temporally variable, and/or variations in the 
residual SWCX emission.

Data from the full field of view were used in this analysis.  
They were conservatively filtered by screening the 
events using the parameters ``FLAG==0'' and ``PATTERN$<$=12'' (only the 
best events selected by position and limiting the CCD pixel pattern to 
legitimate quadruple and lower pixel-number events).  The 
non-cosmic background was modeled using the method of \citet{kea04}, 
which is currently limited to the EPIC MOS data (i.e., excludes data from 
the EPIC PN detector).  The nominal non-cosmic background, which is 
dominated by the quiescent high-energy particle-induced background, 
is modeled using closed filter data scaled by data 
from the unexposed corners of the detectors in the individual observation.  
The scaling is energy dependent and based on the hardness and intensity of 
the corner data (which are temporally variable).  This background modeling 
removes most of the non-cosmic background but unfortunately leaves untouched 
any backgrounds (particle or X-ray) originating in, or passing down the 
telescope tube (i.e., backgrounds which are blocked by the detector mask 
used in the closed-filter data collection).  
Most notably there are strong fluorescent Al~K$\alpha$ 
and Si~K$\alpha$ lines, and the possibility of soft protons at some 
quiescent level.

After background subtraction, the spectra from the full field of view 
of the first three HDF-N pointings and the low state of the fourth were 
reasonably statistically consistent, while the high-state data from the 
fourth pointing showed a strong enhancement at energies less than 
1.4~keV (consistent with the light curve), which was dominated by 
line emission.  The spectra at higher
energies of all four pointings (including both high and low states 
of the fourth) were consistent with each other as expected from their 
average count rates (Table~\ref{tbl:observations}).

\subsubsection{Spectral Fitting}

For the spectral fits of the {\it XMM-Newton} X-ray data we used spectral 
redistribution matrices (RMFs) and effective areas (ARFs) produced by the 
project Standard Analysis Software (SAS) package (Version 5.4.1).
Xspec (Version 11.2, \citet{a01}) was used to fit a model of the cosmic 
X-ray background (CXB), SWCX line emission (a number of discrete lines), 
probable residual soft proton contamination (represented by a power law 
not folded through the instrumental efficiency), and the two instrumental 
lines (Al~K$\alpha$ and Si~K$\alpha$) to 
the data after subtraction of the modeled non-cosmic background.  In order
to constrain the cosmic contribution to the spectrum, we simultaneously 
fit RASS data provided through the HEASARC X-Ray Background 
Tool\footnote{http://xmm.gsfc.nasa.gov/cgi-bin/Tools/xraybg/xraybg.pl},
which uses data from \citet{sea97} to create Xspec compatible spectra
for selected directions on the sky.  The spatial distribution of the CXB 
emission within the observed field, which includes the HDF-N, is relatively 
simple (by design there are no significantly bright point sources or 
extended objects in the field), and can be modeled using the four standard 
components listed in \citet{ks00}: a cooler unabsorbed 
thermal component with $T\sim0.1$~keV representing emission from the Local 
Hot Bubble (e.g., \citet{sea98}), an absorbed cooler thermal component also 
with $T\sim0.1$~keV representing emission from the lower halo, an absorbed 
hotter thermal component with $T\sim0.6$~keV representing Milky Way halo 
or local group emission (e.g., \citet{mea02}), and an absorbed 
power law with a spectral index of 1.46 representing the unresolved 
emission of AGN and other cosmological objects (e.g., \citet{cfg97}).  
The absorption was fixed to the Galactic column density of \hi\ 
($1.5\times10^{20}$~cm$^{-2}$).  

The spectral fitting was done simultaneously for eleven spectra: MOS1 and 
MOS2 data from the four observations with two spectra (high and low states)
for the fourth observation, plus the RASS spectrum.  The CXB model was fit 
to all spectra with no scaling between the different instruments besides
the normalization for solid angle, which was fixed.  The instrumental lines 
and soft-proton power law were fit only to the MOS data. The normalizations 
were allowed to differ between the MOS1 and MOS2 detectors but were 
constrained to be the same for all observations.  The lines representing 
the SWCX were fit only to the MOS data from the high state of the fourth 
observation with the fit parameters constrained to be the same for the 
MOS1 and MOS2 data.

\begin{figure}
\plotone{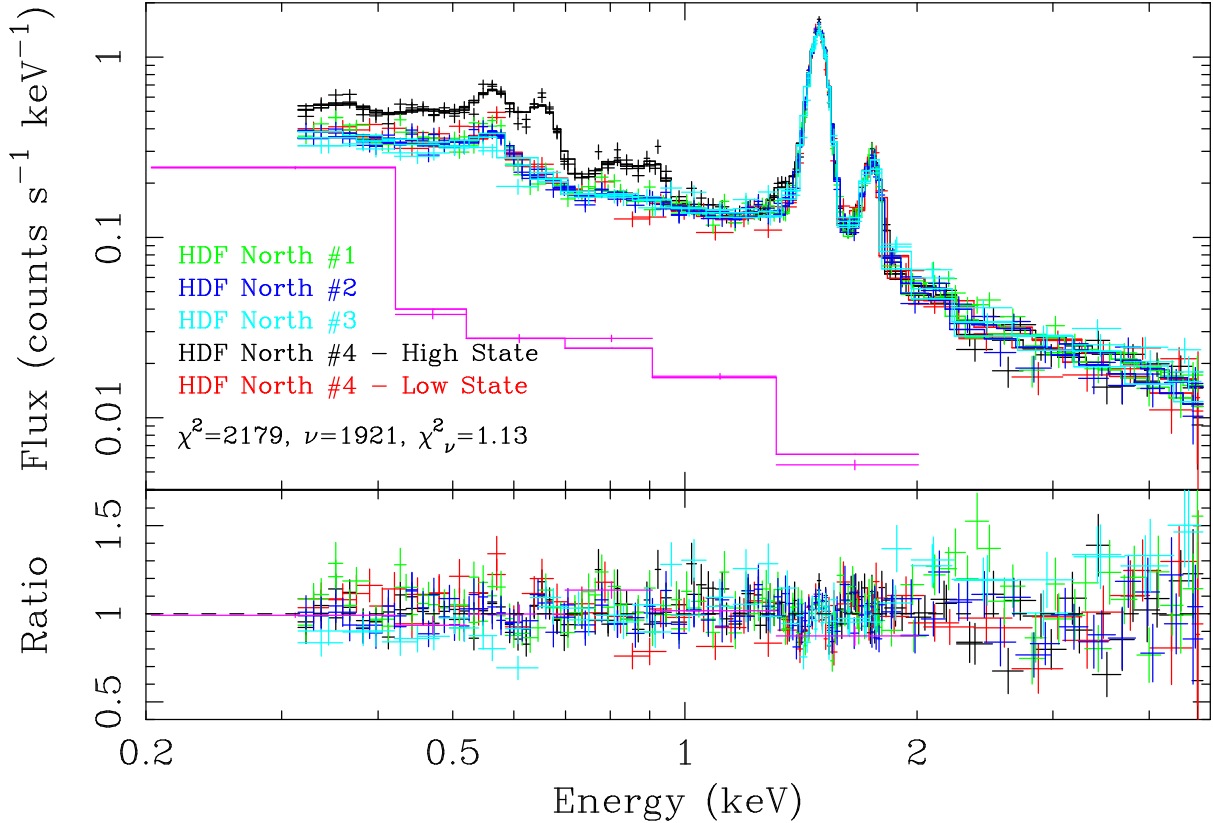}
\caption{Fitted MOS1, MOS2, and {\it ROSAT} All-Sky Survey spectra for 
the four observations of the {\it Hubble} Deep Field North.  The black
curve and data are from Observation \#4 (2001 June 1--2) during the 
higher count rate period (see Figure~\ref{fig:lc}).  The red curve and 
data are from Observation \#4 during the lower count rate period.  The 
green, blue, and light blue curves and data are from observations \#1, 
\#2, and \#3, respectively.  The lower data and curve (purple) are the 
RASS data and model, which have been scaled by a factor of 100 for 
display purposes. The excess of the black spectrum can be modeled 
with \cvi, \ovii, \oviii, \neix, and \mgxi\ emission.  The additional
bump at $\sim0.8$~keV likely consists of \oviii\ lines but may also have 
contributions from \fexvii.  Note the 
absence of a strong \oviii\ bump in the no-excess spectra.  Below the 
instrumental lines ($E<1.4$~keV) the model particle background accounts 
for roughly 25\% of the observed counts, above 2~keV about 40\% of the 
observed counts.
\label{fig:spec}}
\end{figure}

Figure~\ref{fig:spec} shows the results with the excess emission attributed 
to SWCX being clearly visible.  The aluminum and silicon instrumental lines 
are the strong peaks at 1.49~keV and 1.74~keV, respectively.  The SWCX 
emission was modeled by a set of Gaussians whose widths were set to zero 
except for the component at $E\sim0.8$~keV which covers multiple lines from 
\oviii\ and possible 
lines from \fexvii.  The fitted line strengths are listed in 
Table~\ref{tbl:flux}.  Two Gaussian lines with energies less than 
0.5~keV were included in the fits, which can be associated with emission 
from highly ionized carbon (\cvi\ at 0.37~keV and 0.46~keV).  They are 
included in Table~\ref{tbl:flux} with the following caveats.  At energies 
less than 0.5~keV the energy resolution and effective areas of the MOS 
instruments fall off significantly and while the additional lines are 
necessary for reasonable fits ($\Delta\chi^2=39$ when one line is removed,
$\Delta\chi^2=124$ when both lines are removed) the specific energies of the 
lines are not well constrained.  However, there is no degrading of the quality
of the fit when the line energies are fixed at the \cvi\ values, and once the 
energies are fixed fluxes can be measured at better than the 3$\sigma$ level.
Still, lines at other energies may fit the data as well so the attribution
of the emission to \cvi\ is not certain, it is however reasonable (cf. 
\citet{kd00}).  Note, however, that the addition of the \cv\ line at 0.30~keV 
(also listed in \citet{kd00}) did not improve the quality of the fit.

\begin{figure}
\plotone{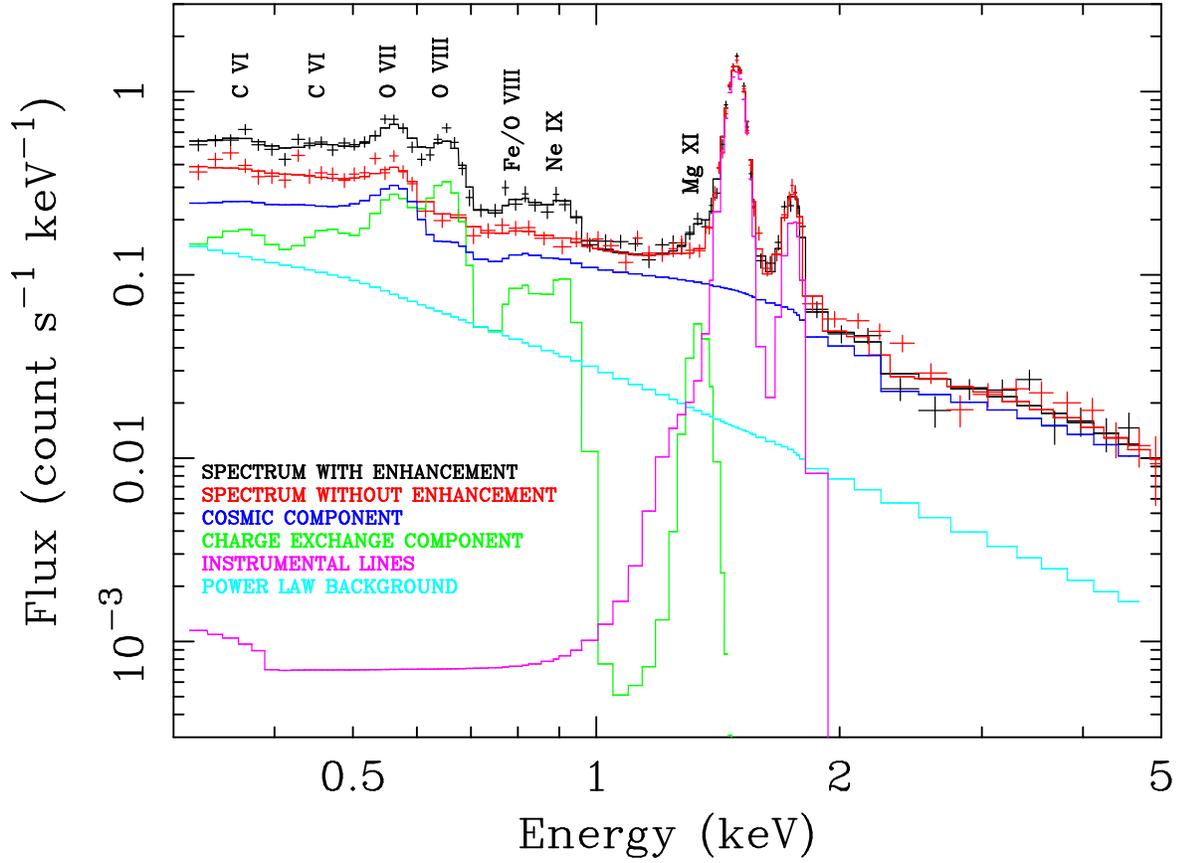}
\caption{Model components of the best spectral fit folded through the
instrumental response.  The MOS1 spectra from HDF-N~\#1 (red) and from the 
SWCX emission period of HDF-N~\#4 (black) are shown along with the 
SWCX, cosmic background, instrumental lines, and power law 
(likely soft proton) background components. \label{fig:fold}}
\end{figure}

The power-law background component, which is assumed to represent 
residual soft proton contamination, was added to bring consistency between 
the RASS and MOS data.  Because this power law is best fit without being 
folded through the MOS effective area, it is not due to X-rays originating 
externally to the satellite as they would be cut off at lower energies by 
the filter.  The effective flux in this component is 
$\sim0.85\times10^{-8}$~ergs~cm$^{-2}$~sr$^{-1}$~s$^{-1}$ while the flux of 
the cosmic background model is 
$7.15\times10^{-8}$~ergs~cm$^{-2}$~sr$^{-1}$~s$^{-1}$ in the 
0.2--5.0~keV band, with 
$3.82\times10^{-8}$~ergs~cm$^{-2}$~sr$^{-1}$~s$^{-1}$ from the 
extragalactic power 
law.  Figure~\ref{fig:fold} shows the MOS1 spectra from the first pointing
and the high state of the fourth pointing along with the various 
components to the fit folded through the instrumental response.

The fit to the data is rather good with $\chi^2_\nu=1.13$ with 1922 
degrees of freedom.  The fitted values for the energies of the lines 
were consistent with the expected values in initial fits, and were 
subsequently fixed (except for the \oviii\ and the possible 
Fe complex at $E\sim0.8$~keV).
In general it is possible to move power between one spectral 
component and another in fits of such complex models as used here.  The
fitted parameters of the cosmic background model are, for example,
strongly correlated.  However, the measurement of the SWCX emission is very 
robust (as indicated by the quoted 90\% probability statistical 
errors in the line fluxes listed in Table~\ref{tbl:flux}) 
as the background is so 
well constrained and the SWCX emission is in a limited number of discrete 
lines.

\subsection{The Solar Wind}
\label{sec:ace}

\subsubsection{Monitoring the Solar Wind}

\begin{figure}
\plotone{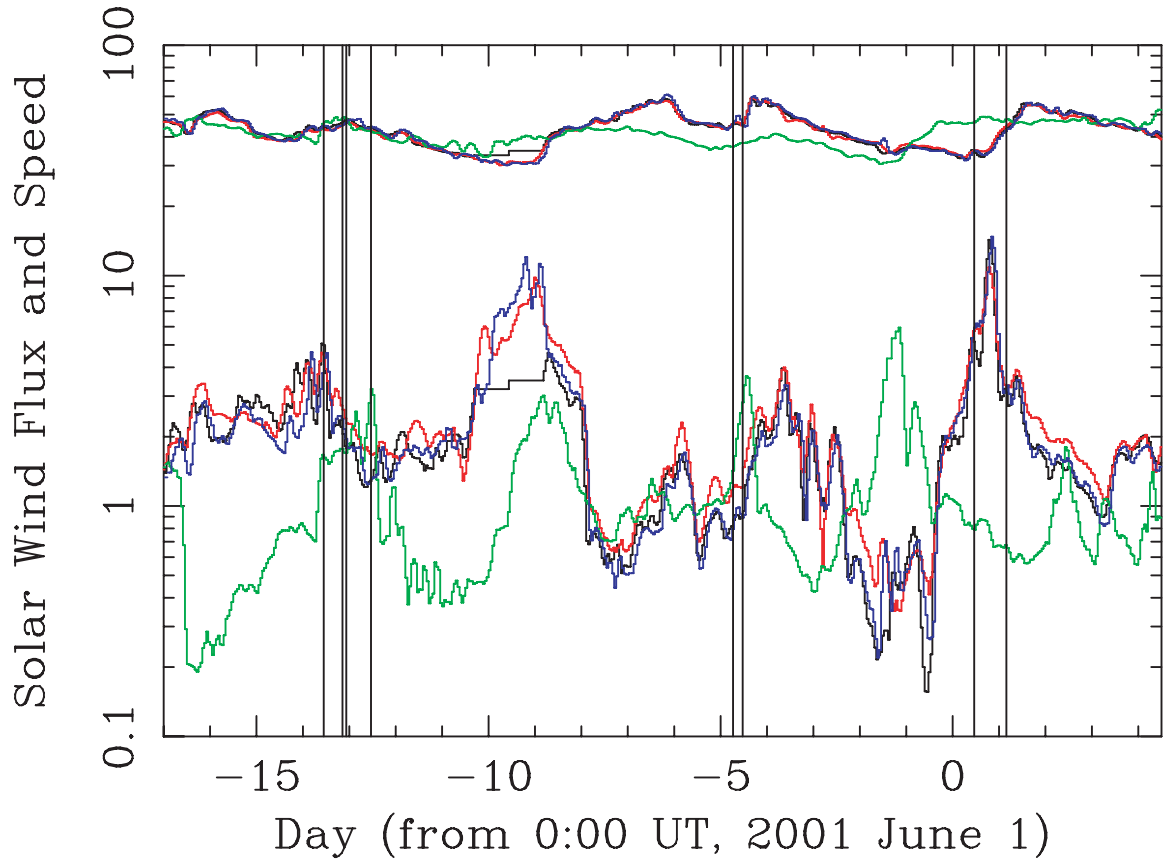}
\caption{Measurement of the solar wind proton speed (upper curves, 
divided by 10 for clarity) and flux (lower curves) by four spacecraft
during the period of the {\it XMM-Newton} HDF-N observations.  The
colors show data from {\it ACE} (black), {\it Wind} (red), {\it SoHO} 
(blue), and {\it Ulysses} (green).  The deviation of the {\it ACE} 
from those of {\it Wind} and {\it SoHO} near day -9 is due to missing 
data.  The vertical bars indicate the time intervals of the four
{\it XMM-Newton} observations.
\label{fig:sw}}
\end{figure}

Solar wind speed and flux data from four spacecraft during the period
of the HDF-N observations are shown in Figure~\ref{fig:sw}.  Three of them, 
{\it ACE}, {\it Wind}, and {\it SoHO} sample the near-Earth environment. 
{\it ACE} and {\it SoHO} have been, for the most part, located in halo 
orbits around the Sun-Earth libration point (L1, $\sim1.5\times10^6$~km, 
235~$R_E$ from Earth).  {\it Wind} has led a more peripatetic life and at 
the time of these measurements was no longer in an orbit around L1.
{\it ACE} was launched on 1997 August 25, 
{\it SoHO} was launched on 1995 December 2, and {\it Wind} was launched 
1994 November 1.  The three spacecraft function, in part, as upstream 
monitors of the solar wind, the hot ($\sim10^6$~K) plasma continually 
flowing from the Sun, and provide almost continual measurements of the 
solar wind density, speed, and magnetic field, among many other parameters.
The fourth spacecraft, {\it Ulysses}, follows a high-inclination 
($i\sim79^\circ$), elliptical ($e\sim0.6$ with a semi-major axis of 
$\sim3.4$~AU) orbit around the Sun.  {\it Ulysses} was launched 1990 
October 6 and reached its roughly polar orbit after a fly-by of Jupiter.

At the time of these measurements, Earth was at $\sim1.01$~AU, $0.0^\circ$ 
ecliptic latitude, and $\sim251^\circ$ ecliptic longitude.  The solar wind 
speed and flux data from the three near-Earth spacecraft are in reasonably 
good agreement showing the same structure (Figure~\ref{fig:sw}).  The major 
enhancements in the 
flux at all three spacecraft are roughly contemporaneous while discrepancies 
in the fluxes, although typically small but sometimes as much as a factor of 
two, may be in part due to the sometimes large separation ($>200~R_E$) between 
the spacecraft perpendicular to the Sun-Earth line.  (Note that the deviation 
of {\it ACE} near day --9 in Figure~\ref{fig:sw} is due to missing data.)  
On the other hand {\it Ulysses}, which is in a polar orbit around the Sun, 
at this time was $\sim1.34$~AU from the Sun with an ecliptic latitude of 
$\sim5.6^\circ$ and an ecliptic longitude of about $\sim339^\circ$.
This location is farther from the Sun than the other spacecraft and 
$\sim90^\circ$ ahead of the Earth.  Thus, the Ulysses flux in 
Figure~\ref{fig:sw} is only qualitatively similar to that observed near 
Earth and shifted in time. These differences reflect the large scale
structure of the interplanetary medium at this time.  In addition, the 
{\it Ulysses} direction was $\sim110^\circ$ away from the HDF-N, and so 
is not particularly relevant for the SWCX observation, but the comparison 
of the data provides an illustration of the variation of the solar wind 
from a uniform shell-like expansion.

\subsubsection{The Solar Wind Enhancement}

Although there exists considerable variability, solar wind 
speeds are typically about 450~km~s$^{-1}$ (about 1 keV/nucleon)
with fluxes at 1~AU of about $3\times10^8$~cm$^{-2}$~s$^{-1}$ \citep{rea96}.
The solar wind by is about 95\% protons and 4\% helium nuclei by number. 
High charge state heavy ions (i.e., other than protons and He$^{+2}$) comprise 
less than 1\% of the solar wind by number. These ions include 
carbon, dominated by charge states C$^{+5}$ and C$^{+6}$, 
oxygen, dominated by charge states O$^{+6}$ and O$^{+7}$, neon 
dominated by charge state Ne$^{+8}$, and iron with a wide range 
of charge states sometimes as high as Fe$^{+17}$ \citep{gla99,col96}. 
The solar wind source ion for the \cvi\ lines in the X-ray spectrum 
is C$^{+6}$ and the source ions for the \ovii\ and \oviii\ lines 
are O$^{+7}$ and O$^{+8}$, respectively.  Although typically 
the dominant charge states of solar wind oxygen are +6 and +7, under 
certain conditions, e.g., in the CME-related solar wind (coronal mass ejection), 
O$^{+8}$ can be comparable to and even dominate the O$^{+7}$ \citep{g97}.  
Ne$^{+9}$ and Mg$^{+11}$, the solar wind source ions for 
\neix\ and \mgxi\ SWCX emission, have also been observed in the solar 
wind (A. B. Galvin, private communication).  \oviii\ and \neix\ lines
have been observed in cometary spectra \citep{kea03}

During the period of the 2001 June 1-2 {\it XMM-Newton} HDF-N pointing, 
the {\it ACE} Level~2\footnote{{\tt http://www.srl.caltech.edu/ACE/ASC/}} 
data show that the O$^{+7}$/O$^{+6}$ ratio was highly variable 
(Figure~\ref{fig:lc}) compared to the first three pointings.  The average 
ratio was enhanced by over a factor of two during the period when the SWCX 
emission was detected, and dropped by nearly an order of 
magnitude at the end of the pointing concurrent with the reduction 
in solar wind flux (Table~\ref{tbl:ace}).  This large increase in the 
relative abundance of the higher ionization state during the SWCX period 
might suggest that the relative abundances of the ionization states 
of O$^{+8}$, Ne$^{+9}$, and Mg$^{+11}$ may also be significantly enhanced.  
Thus observation of SWCX emission from these species
would be reasonable.  However, preliminary results from {\it ACE} indicate
that the O$^{+8}$/O$^{+7}$ ratio was actually relatively low at $\sim0.05$
during the period of SWCX emission (Zurbuchen \& Raines, private 
communication; the analysis continues).  The picture becomes even more murky
as the preliminary results also show the O$^{+8}$/O$^{+7}$ ratio increasing 
by a factor of 20 to 30 just as the SWCX emission and solar wind flux 
enhancement were cutting off.  In addition, the 
(0.72--0.61~keV)/(0.52--0.61~keV) count rate ratio, which is a measure of
the relative emission of \oviii\ and \ovii, was consistent with a
constant value ($\chi^2=25.8$ for 22 degrees of freedom) over the period
of the SWCX emission.  This is in spite of the variation in solar wind 
ionization state distributions.

The temporal variation of the X-ray data requires that the enhancement 
originates locally, at least in the solar system if not in the 
satellite/near-Earth environment.  The presence of X-ray emission 
lines corresponding to the highly ionized charge states of oxygen, 
neon, and magnesium eliminates an origin internal to the satellite, and 
strongly implicates SWCX as the source of the emission enhancement. 
Furthermore, data from the Low Energy Neutral Atom {\it LENA} imager 
on the {\it IMAGE} spacecraft inside the magnetosphere during this 
event show an enhancement which suggests a neutral solar wind resulting 
from SWCX \citep{col01}.

\section{Discussion}
\label{sec:discussion}

\subsection{Observation Geometry}
\label{sec:geometry}

\begin{figure}
\epsscale{0.8}
\plotone{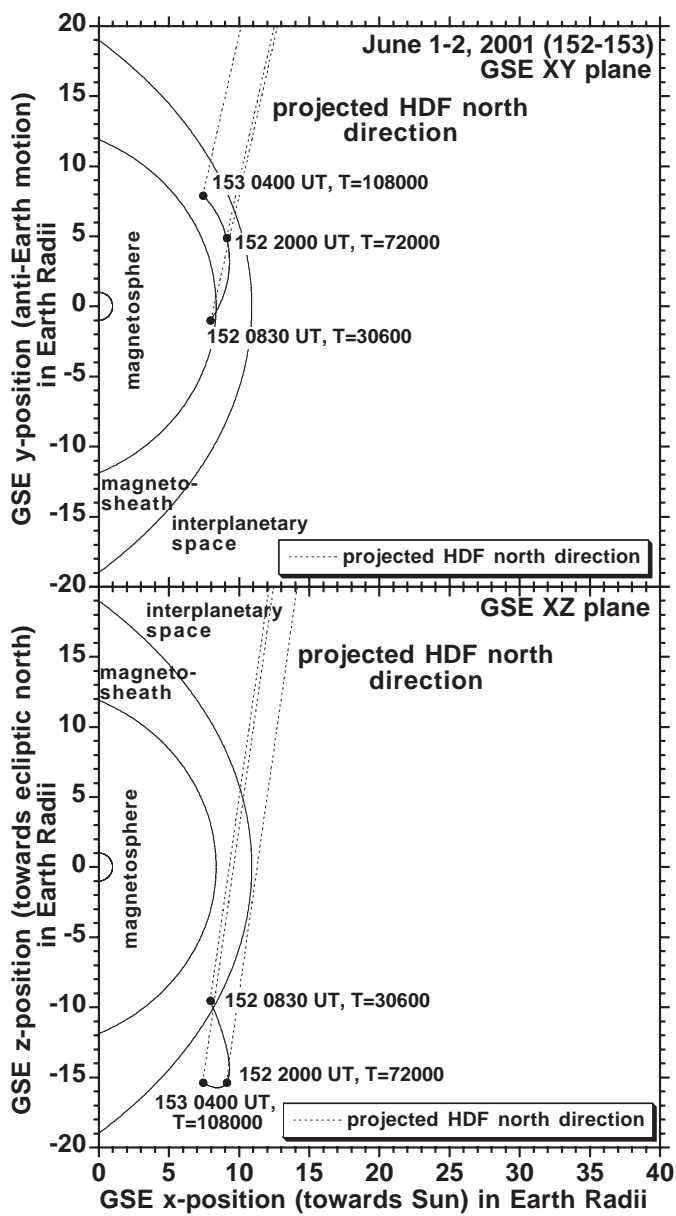}
\caption{The geometry of the 2001 June 1 (fourth HDF-N) 
observation.  The upper
panel shows the observation line of sight projected onto the plane 
of the ecliptic where the positive X-axis points toward the Sun and 
the positive Y-axis is opposite to Earth's velocity vector.  The 
bottom panel shows the observation line of sight projected onto the 
vertical plane where the positive Z-axis is toward the north ecliptic 
pole. Note that the line of sight is slightly sunward and lies in all 
cases outside of the magnetosphere.  However, also note that there is 
some uncertainty in the exact location in the magnetosheath and 
magnetosphere boundaries.
\label{fig:geometry}}
\end{figure}

The geometry of the 2001 June 1--2 observation with respect to 
the near-Earth environment is shown in Figure~\ref{fig:geometry}.  
The satellite is on the Sun side of the Earth with an angle of 
$\sim83^\circ$ between the Sun and look directions.  (Due to 
satellite constraints, the look direction of {\it XMM-Newton}
must lie within $20^\circ$ of perpendicular to the Earth-Sun 
line.)  The satellite and the observation line of 
sight are outside of Earth's magnetosphere, based on the 
magnetospheric model of \citet{pr96} (using a ram pressure of 
7.4~nPa and a northward $B_z$).  However, the line of sight 
may possibly pass through the magnetosheath relatively near the 
sub-solar point.  Such an observation trajectory would sample 
the region of strongest production of X-rays by SWCX with 
exospheric material \citep{rc03a}.  On the other hand, the 
trajectory may be entirely outside of the magnetosheath 
and therefore miss much of that emission.
Because of this uncertain geometry, the SWCX emission can be 
from either exospheric material in Earth's magnetosheath or from 
interstellar neutrals in the heliosphere, or more likely both (at 
least to some level).  The detailed analysis of this geometry will 
be the subject of a following paper.

On a more global scale (with perhaps no particular relevance 
to this observation), during the time period of the 
2001 June 1--2 observation the location of Earth in its
orbit placed it almost directly upstream from the Sun relative to 
the flow of interstellar material through the solar system.
(Earth is in the nominal upstream direction about 4~June.)

\subsection{The SWCX Emission Spectrum}

From Table~\ref{tbl:flux} and Figure~\ref{fig:spec} it is clear that the 
excess emission can be well characterized by a limited number of highly 
significant lines with energies consistent with those expected from SWCX.  
With certain assumptions it is possible to use the observed X-ray line 
ratios to derive the concurrent ion ratios.  

The fitted \oviii(0.65\&0.81~keV)/\ovii(0.56~keV) line ratio from 
Table~\ref{tbl:flux} is 1.09.  
However, from \citet{kea03}, the $3^3P\rightarrow1^1S$ and (at the 2\% level) 
\ovii~$3^1P\rightarrow1^1S$ lines contaminate the \oviii\ line at $\sim0.65$~keV, 
and contribute $\sim0.58$~cm$^{-2}$~sr$^{-1}$~s$^{-1}$ (using a factor of 
0.079, derived from the relative intensities of the relevant lines in their 
Table 1, to scale from the \ovii\ 0.56~keV flux).  Correcting for this 
contribution, and assuming that the 0.81~keV emission is all from \oviii\ 
(see below), the \oviii(0.65\&0.81~keV)/\ovii(0.56\&0.67~keV) emission ratio 
is then $\sim0.94$
(Table~\ref{tbl:oxygen}).  To convert the emission ratio to an ion number
ratio we use the SWCX cross sections in \citet{wea04}. 
The ratios between the hydrogen and helium SWCX cross sections for 
O$^{+7}$ and O$^{+8}$ are roughly the same, so we use the 
O$^{+8}$/O$^{+7}$ hydrogen cross section ratio of 1.66 to scale the 
flux ratio (this assumes that the scattering is optically thin and 
hydrogen is the dominant SWCX target neutral).  This implies a 
solar wind O$^{+8}$/O$^{+7}$ abundance ratio of $0.57\pm0.07$, which 
is somewhat higher than the more typical value of 0.35 
(e.g., \citet{sc00}), but not unreasonable, as would be expected
considering the enhancement of the O$^{+7}$/O$^{+6}$ ratio. However,
as noted above, the preliminary O$^{+8}$/O$^{+7}$ ratio from {\it ACE}
for the SWCX period is $\sim0.05$, a significant discrepancy which we
will revisit in a subsequent paper when the O$^{+8}$/O$^{+7}$ ratio
data are more robust.

The energy and non-zero width for the fitted line at $E\sim0.8$~keV is 
consistent with the expected \oviii\ Ly$\beta$ through Ly$\epsilon$ 
emission (\citet{wea04}).  However, this is also the energy of possible 
\fexvii\ emission.  The ratio of the line strengths, 
Ly$\alpha$/Ly($\beta-\epsilon$), predicted by \citet{wea04} is 2.75
(with a $\sim30\%$ uncertianty), while our measured ratio is $3.9\pm0.8$.  
The two values are therefore in reasonably good 
agreement but the result does not rule out a contribution from \fexvii.
It does suggest that any Fe contribution to the observed flux is 
relatively limited.

\subsection{Contemplation of the Temporal Variation}

As measured by {\it ACE}, the solar wind during the 2001 June 1--2 
period was in its ``slow'' state but with with fairly uncharacteristic 
ion abundances and flux during the enhancement period.    
At solar maximum, the solar wind at 1 AU shows a complex structure most 
likely resulting from the appearance of equatorial coronal holes. In 
contrast, at solar minimum, there tends to be a two state speed 
structure observed in the solar wind with fewer but well-formed 
transients such as coronal mass ejections and corotating interaction 
regions.  \citet{smea03} provide a recent review of the complex 
variation of the solar wind during the solar cycle.


Solar wind fluxes during the first three {\it XMM-Newton} HDF-N observations 
were close to nominal levels (see Figure~\ref{fig:sw}).  However, during 
the observation on 2001 June 1-2, the solar wind flux observed by the ACE 
spacecraft was  enhanced (as shown in Figures~\ref{fig:lc} and \ref{fig:sw}) 
by a factor of five over its typical value of  
$\sim3\times10^8$~cm$^{-2}$~s$^{-1}$.  The average O$^{+7}$/O$^{+6}$ ionization 
state ratios for the observation (Table~\ref{tbl:ace}) were also 
significantly enhanced. 
This flux enhancement was relatively localized; it lasted 5-6 hours at a solar 
wind speed of $\sim350$~km~s$^{-1}$ implying a width of about 0.05~AU in 
the direction toward the Sun.  The X-ray flux decrease at $T\sim81000$ 
(2001 June 1, 22:30 UT) and the sharp drop in both the solar wind flux 
and O$^{+7}$/O$^{+6}$ ratio propagated 
to Earth from the L1 point (a delay of $\sim4000$~s) are approximately 
contemporaneous, which suggests the X-ray flux is 
responding primarily to local solar wind conditions.  

However, comparison of the solar wind flux and ionization state 
variation and X-ray light curves raises two
critical questions.  First, if the SWCX emission enhancement observed by 
{\it XMM-Newton} is due to a local phenomena, why is the large variation 
in solar wind flux not reflected in the X-ray intensity, or in a variation 
of the relative intensity of the \ovii\ and \oviii\ emission?  Over the course 
of 20~ks (roughly six hours) the solar wind flux increases by a factor of
six while the X-ray light curve is essentially constant (Figure~\ref{fig:lc}).
If the SWCX emission was due solely to 
interactions with material from Earth's exosphere, then the X-ray light curve 
should be closely linked to the solar wind flux measured by {\it ACE} at the 
L1 point.  \citet{rc03a} point out that the geometry of the magnetosheath is
strongly affected by the density and velocity of the solar wind.  With the 
observation geometry considered by \citet{rc03a}, i.e., looking out through 
the flanks of the magnetosheath, there is some buffering of the X-ray emission
which smooths out the effects of strong solar wind enhancements.  However, 
with the geometry of this observation where the line of sight possibly runs
tangentially through their region of maximum X-ray emission near the 
magnetosheath sub-solar point, the situation is much more confused, and 
perhaps more likely to produce variations rather than to smooth out the 
SWCX flux.    

\begin{figure}
\plotone{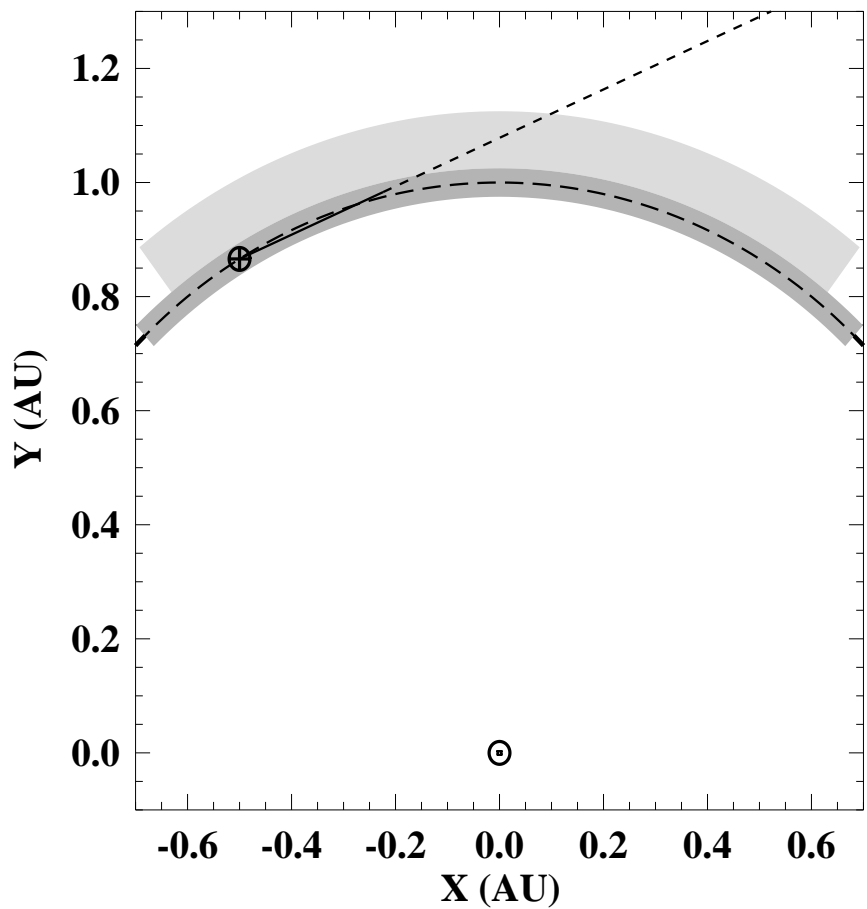}
\caption{A cartoon of the solar system geometry for the observation of
charge exchange from the density enhancement in the solar wind.  The X,Y 
plane is defined by the Earth-Sun line and the look direction.  The strongest 
part of the enhancement, shown as the darkest shading, lasts about a quarter 
of a day, which with the typical velocity of the solar wind (about a quarter 
of an AU per day) extends about 5\% of an AU.  The solar wind density was 
enhanced by about a factor of three (shown by the lighter shading) for 
more than a half a day preceding the main enhancement.  The solid line from
the Earth shows the look direction, and shows how the solar wind enhancement
is sampled with the assumption of a spherically symmetric propagation.
\label{fig:geometry2}}
\end{figure}

Given the possible 
difficulty in turning the large variation of the solar wind flux into a
relatively constant X-ray light curve, a possible solution might be to
distribute the emission along the line of sight through the heliosphere.
This would smooth out temporal variations originating with the solar 
wind.  However, this brings up the second question.  If the emission is 
distributed through the heliosphere why do the solar wind and X-ray 
enhancements cut off at nearly the same time?   Note again the large 
differential variation in the solar wind flux as measured by {\it ACE} 
and {\it Ulysses}, up to a factor of 20, during SWCX observation.  
Figure~\ref{fig:geometry2} 
illustrates the observation geometry on a larger scale.  
Since the pointing direction is $\sim83^\circ$ 
from the Earth-Sun line, the observation samples a relatively long path length 
through the solar wind enhancement with a heliocentric distance similar to that 
of the Earth.  If the solar wind enhancement is a spherically symmetric shell 
moving outward there should still be a strong enhancement in the SWCX emission 
roughly concurrent with the {\it ACE} measurement. However, even introducing 
large variations in the density profile and arrival time of the solar wind 
enhancement as a function of direction from the Sun will not in general smooth 
out the X-ray light curve significantly.  In addition, once the position along the
line of sight is greater than 1~AU from the Sun (the short dashed part of the 
line of sight in Figure~\ref{fig:geometry2}), it is sampling the solar wind 
enhancement after it would have passed Earth, and therefore the X-ray count
rate should fall off after the fall-off of the solar wind 
enhancement as measured at Earth.  One might argue that such a fall-off 
could slow down and therefore not be obvious in the relatively limited lower
flux part of the 0.52--0.75~keV light curve.  However, the absence of 
\oviii\ emission during that period indicates that the SWCX emission really
has cut off.

The quandary is therefore: 1) If the X-ray emission is from the magnetosheath 
why is it so constant? 2) if the X-ray emission is distributed through the 
heliosphere why does the emission fall off concurrently with the end of the 
solar wind enhancement, and why (again) is the SWCX emission light curve so 
constant?

\subsection{Comparison with RASS LTEs}

The episode of SWCX emission reported here lasted for at least 38~ks 
($\sim10.6$ hours) which would have extended over seven {\it ROSAT} 
orbits.  This is sufficiently long enough to be classified as a LTE 
in the RASS processing.  However, this SWCX episode has considerably 
more \tqkev\ flux than LTEs observed during the RASS.  The fluxes 
listed in Table~\ref{tbl:flux} would have produced a {\it ROSAT} PSPC 
count rate of $\sim210\times10^{-6}$~counts~arcmin$^{-2}$~s$^{-1}$ in 
the R45 ($3\over4$~keV) band, only slightly less than twice the typical 
intensity of the cosmic background at high Galactic latitudes.  
Most LTEs were dominated by emission in the R12 ($1\over4$~keV) band and 
the brightest $3\over4$~keV band LTE enhancements were only roughly half 
of the intensity of this detection. 

\section{Conclusions}
\label{sec:conclusions}

The data from the {\it XMM-Newton} observation of the {\it Hubble} Deep 
Field North show clear evidence for a time-variable component of the 
X-ray background that is most reasonably
attributed to charge exchange emission between the highly ionized solar
wind and exospheric or interplanetary neutrals.  The SWCX emission is 
dominated by the lines from \cvi, \ovii, \oviii, \neix, and \mgxi, and 
possibly lines from 
highly ionized iron (e.g., \fexvii).  The emission is concurrent with the 
passage of a strong enhancement in the solar wind observed by the {\it ACE} 
satellite, and which is also strongly enhanced in its O$^{+7}$/O$^{+6}$ 
ratio, and possibly other highly ionized species (although preliminary results
for the O$^{+8}$/O$^{+7}$ ratio indicate otherwise).  
However, the light curve of the X-ray enhancement is relatively
constant while the flux of the solar wind enhancement varies
considerably.

The observation of SWCX emission allows the monitoring of interactions 
between the solar wind and solar system and/or interstellar neutrals without
the need for {\it in situ} measurements, albeit with the uncertainty of just
where along the line of sight the emission arises.  However, in certain 
circumstances the distance ambiguity can be resolved allowing the detailed 
study of the related phenomena.  Two such situations are mentioned above: the
case of SWCX emission from comets and exospheric material in the {\it Chandra} 
dark moon observation.  The ability to remotely sense the solar wind and its 
interactions can also be used when the emission is expected to have a
predictable variation, such as an X-ray scan across the magnetopause sub-solar
point \citep{rc03a}.  Also, for example, {\it ROSAT} data from the All-Sky 
Survey suggest that the downstream helium focusing cone can be imaged in 
X-rays (e.g., \citet{cea03}).

The charge exchange line emission can provide a significant 
contaminating background 
to observations of more distant objects (by any X-ray observatory with 
sensitivity at energies less than 1.5~keV) which use those lines for 
diagnostics.  However, the correlation of the X-ray enhancement with 
the solar wind density enhancement suggests a diagnostic.  For those 
observations at risk because of such contamination, e.g., observations of 
sources which cover the entire field of view and have thermal spectra, 
the data from solar wind monitoring observatories such as {\it ACE}, 
{\it Wind}, and {\it SoHO} should 
be used for screening.  Should the SWCX emission be primarily exospheric
in origin (e.g., from the magnetosheath), then the specific geometry of
the observation should be examined as well.  An observation where the 
observatory is outside of the magnetosheath and looking away from Earth 
would clearly have significantly less exospheric SWCX emission than the 
case of the emission reported in this paper where the observation line 
of sight may be passing tangentially through the magnetopause near the 
sub-solar point.

While SWCX emission is likely responsible for the LTEs observed during
the {\it ROSAT} All-Sky Survey, the SWCX episode presented here was 
significantly brighter at \tqkev\ than what was typically observed
(a factor of two brighter than the most intense \tqkev\ RASS LTE).  
However, one important issue to note is that none of the X-ray emission 
lines discussed here (or discussed in \citet{wea04}) contribute 
significantly to the \oqkev\ band, where most of the LTE emission was 
observed.  Also of note is that the geometry of this observation relative 
to Earth's geomagnetic environment is considerably different from that of 
the RASS. In this case we may be looking tangentially through the 
magnetosheath near the sub-solar point while {\it ROSAT} with its low 
circular orbit effectively looked radially outward through the flanks of 
the magnetosheath.

\acknowledgments

We are grateful to the {\it ACE} SWEPAM and SWICS/SWIMS instrument teams 
and the {\it ACE} Science Center for providing the {\it ACE} data, to 
{\it Wind}/SWE team for providing the {\it Wind} data, to D. J. McComas, 
M. Lancaster, and C. Tranquille for providing the {\it Ulysses}/SWOOPS, 
and to {\it SoHO} team for providing the CELIAS/MTOF PM data.
We would also like to thank M. Bzowski, S. Christon, T. Cravens, A. Galvin, 
C. Lisse, B. Pilkerton, J. Raines, 
I. Robertson, D. Simpson, A. Szabo, B. Wargelin, and T. Zurbuchen for useful 
discussions.  We also thank B. Wargelin for providing a copy of 
\citet{wea04} to us before submission and the referee for their comments.  
Finally, we would like to thank the referee for their comments.
This work is based on observations 
obtained with {\it XMM-Newton}, an ESA science mission with instruments and 
contributions directly funded by ESA Member States and the USA (NASA).  The 
data were provided through the HEASARC {\it XMM-Newton} archive at NASA/GSFC.

\clearpage

\clearpage 

\clearpage 

\clearpage 

\clearpage 

\clearpage 

\clearpage

\begin{deluxetable}{ccccccc}
\tablecaption{{\it XMM-Newton} {\it Hubble} Deep Field North Observations.
\label{tbl:observations}}
\footnotesize
\tablewidth{0pt}
\tablehead{
\colhead{Observation ID} & \colhead{Start Date and Time}   
& \colhead{Exposure}   & \colhead{Useful Exp\tablenotemark{a}} 
& \colhead{Count Rate\tablenotemark{b}} & \colhead{Count Rate\tablenotemark{b}} \\
 & & (ks) & (ks) & \colhead{0.52--0.75 keV}  & \colhead{2.0--8.0 keV}
}
\startdata
0111550101 & 2001-05-18T08:46:57 & 46.4 & 39.5 & $0.078\pm0.002$ & $0.413\pm0.004$ \\
0111550201 & 2001-05-18T22:17:34 & 46.6 & 35.1 & $0.080\pm0.002$ & $0.407\pm0.005$ \\
0111550301 & 2001-05-27T06:15:22 & 52.6 & 13.8 & $0.074\pm0.003$ & $0.428\pm0.007$ \\
0111550401\tablenotemark{c} & 2001-06-01T08:16:36 & 95.4 & 38.1 
& $0.126\pm0.002$ & $0.405\pm0.003$ \\
0111550401\tablenotemark{d} & -- & -- & 17.2 
& $0.084\pm0.002$ & $0.400\pm0.005$ \\
\enddata
\tablenotetext{a}{Exposure remaining after time periods affected by obvious 
soft-proton flaring have been excluded.}
\tablenotetext{b}{Average EPIC MOS1 count rate for the flare-free 
periods in counts~s$^{-1}$.  The count rates include the contributions of the 
quiescent particle background.}
\tablenotetext{c}{Charge exchange emission interval.}
\tablenotetext{d}{Quiescent emission interval.}
\end{deluxetable}

\begin{deluxetable}{ccccccc}
\tablecaption{Fitted heliospheric emission line fluxes. \label{tbl:flux}}
\tablewidth{0pt}
\tablehead{
\colhead{Line} & \colhead{Energy} & \colhead{Photon Flux\tablenotemark{a}}         
& \colhead{Energy Flux} \\  
& \colhead{keV}    & \colhead{(cm$^{-2}$~sr$^{-1}$~s$^{-1}$)}   
& \colhead{($10^{-9}$~ergs~cm$^{-2}$~sr$^{-1}$~s$^{-1}$)} 
}
\startdata
\cvi\tablenotemark{b}    & 0.37 & $9.0\pm2.5$   & 5.3 \\
\cvi\tablenotemark{b}    & 0.46 & $2.9\pm0.8$   & 2.1 \\
\ovii\tablenotemark{c}   & 0.56 & $7.39\pm0.79$ & 6.66 \\
\oviii\tablenotemark{d}  & 0.65 & $6.54\pm0.34$ & 6.84 \\
\oviii\tablenotemark{e}  & 0.81 & $1.52\pm0.28$ & 1.96 \\
\neix\                   & 0.91 & $0.85\pm0.22$ & 1.24 \\
\mgxi\                   & 1.34 & $0.40\pm0.08$ & 0.85 \\
\enddata
\tablenotetext{a}{The quoted errors are the 90\% confidence limits.}
\tablenotetext{b}{The energies of the \cvi\ lines were fixed.  See the text
for caveats concerning the robustness of attributing the flux to \cvi.}
\tablenotetext{c}{The \ovii\ ``line'' at 0.56 keV is a complex made up of
the $2^3S\rightarrow1^1S$, $2^3P\rightarrow1^1S$, and $2^1P\rightarrow1^1S$
transitions.}
\tablenotetext{d}{\oviii\ $2p\rightarrow1s$ is the dominant transition, but 
there are contributions from \ovii\ $3^3P\rightarrow1^1S$ and, to a very 
limited extent, the $3^1P\rightarrow1^1S$ transitions.}
\tablenotetext{e}{Includes multiple \oviii\ Lyman lines and possible 
\fexvii\ emission.}
\end{deluxetable}

\begin{deluxetable}{ccccccc}
\tablecaption{Average {\it ACE} solar wind values during the 
{\it XMM-Newton} observations.
\label{tbl:ace}}
\footnotesize
\tablewidth{0pt}
\tablehead{
\colhead{Observation ID} & \colhead{H$^+$ Speed\tablenotemark{a}}  
& \colhead{H$^+$ Density\tablenotemark{a}}   
& \colhead{H$^+$ Flux} & \colhead{O$^{+7}$/O$^{+6}$ Ratio\tablenotemark{b}} \\
                         & \colhead{(km s$^{-1}$)} & \colhead{(cm$^{-3}$)} 
& \colhead{($10^8$~cm$^{-2}$ s$^{-1}$)}
}
\startdata
0111550101 & 447 &  6.4 & 2.9 & 0.46 \\
0111550201 & 451 &  3.3 & 1.5 & 0.43 \\
0111550301 & 457 &  2.0 & 0.9 & 0.41 \\
0111550401\tablenotemark{c} & 345 & 23.3 & 8.0 & 0.99 \\
0111550401\tablenotemark{d} & 418 &  7.6 & 3.2 & 0.15 \\
\enddata
\tablenotetext{a}{{\it ACE} SWEPAM data.}
\tablenotetext{b}{{\it ACE} SWICS-SWIMS data.}
\tablenotetext{c}{Charge exchange emission interval.}
\tablenotetext{d}{Quiescent emission interval.}
\end{deluxetable}

\begin{deluxetable}{ccccc}
\tablecaption{Total\tablenotemark{a}\ \ SWCX oxygen ion 
emission fluxes. \label{tbl:oxygen}}
\tablewidth{0pt}
\tablehead{
\colhead{SW Ion} & Energy & \colhead{Photon Flux} \\  
& \colhead{(keV)}    & \colhead{(cm$^{-2}$~sr$^{-1}$~s$^{-1}$)}
}
\startdata
O$^{+7}$  & 0.56 \& 0.65 & $7.97\pm0.85$  \\
O$^{+8}$  & 0.65 \& 0.81 & $7.48\pm0.44$  \\
\enddata
\tablenotetext{a}{The O$^{+7}$ flux includes the modeled contributions
from the \ovii~$3^1P\rightarrow1^1S$ and $3^3P\rightarrow1^1S$ transitions. 
The O$^{+8}$ flux excludes the modeled contributions from the 
\ovii~$3^1P\rightarrow1^1S$ and $3^3P\rightarrow1^1S$ transitions but adds
all of the emission at 0.81 keV assuming that it is from 
\oviii~Ly(${\beta-\epsilon}$).}
\end{deluxetable}

\begin{deluxetable}{cccccc}
\tablecaption{Heliospheric Emission and Abundance Ratios. 
\label{tbl:ratios}}
\tablewidth{0pt}
\tablehead{
\colhead{SW Ion Ratio}  & \colhead{X-ray Flux}  
& Implied Ion & \colhead{{\it ACE} Ion Ratio}
& \colhead{Typical Ion Ratio\tablenotemark{b}} \\
& \colhead{Ratio\tablenotemark{a}} & \colhead{Ratio} &  \colhead{(2001 June 1)} 
& \colhead{(Slow Wind)}  
} 
\startdata
O$^{+7}$/O$^{+6}$        & --             & --             & 0.99 
      & 0.27 \\ 
O$^{+8}$/O$^{+7}$        & $0.94\pm0.11$  & $0.57\pm0.07$  & -- 
      & 0.35 \\ 
Ne$^{+9}$/O$^{+7}$\tablenotemark{c}  & $0.11\pm0.03$  & --             & -- 
      & --  \\
Mg$^{+11}$/O$^{+7}$      & $0.05\pm0.01$  & --             & -- 
      & -- \\
\enddata
\tablenotetext{a}{The O$^{+7}$ flux includes the modeled contributions
from the \ovii~$3^1P\rightarrow1^1S$ and $3^3P\rightarrow1^1S$ transitions. 
The O$^{+8}$ flux excludes the modeled contributions from the 
\ovii~$3^1P\rightarrow1^1S$ and $3^3P\rightarrow1^1S$ transitions but adds
all of the emission at 0.81 keV assuming that it is from 
\oviii~Ly(${\beta-\epsilon}$).}
\tablenotetext{b}{From \citet{sc00}.}
\tablenotetext{c}{Ne$^{+9}$ has been detected in the solar wind \citep{g97}.}
\end{deluxetable}


\begin{thebibliography}{}
\bibitem[Arnaud(2001)]{a01} Arnaud, K. A. 2001, ASP Conference Proceedings, 
      238, 415
\bibitem[Chen, Fabian, \& Gendreau(1997)]{cfg97} Chen, L.-W., Fabian, A. C.,
      \& Gendreau, K. C. 1997, MNRAS, 285, 449
\bibitem[Cox(1998)]{cox98} Cox, D. P. 1998, Lecture Notes in Physics,
      (Berlin:Springer Verlag), 506, 121
\bibitem[Collier et al.(1996)]{col96} Collier, M. R., Hamilton, D. C., 
      Gloeckler, G., Bochsler, P., Sheldon, \& R. B. 1996, \grl, 23, 1191
\bibitem[Collier et al.(2001)]{col01} Collier et al. 2001, \jgr, 106, 24893
\bibitem[Collier et al.(2003)]{cea03} Collier, M. R., Snowden, S., 
      Moore, T. E., Simpson, D., Pilkerton, B., Fuselier, S., \& Wurz, P.
      2003, Eos Trans. AGU, 84(46), Fall Meet. Suppl., 
      Abstract SH11A-1123
\bibitem[Cravens(1997)]{cra97} Cravens, T. E. 1997, \grl, 24, 105 
\bibitem[Cravens(2000)]{cra00} Cravens, T. E. 2000, \apj, 532, L153 
\bibitem[Freyberg(1994)]{frey94} Freyberg, M. J., 1994, Ph.D. Thesis, 
      Teschnische Universit\"at M\"unchen
\bibitem[Galvin(1997)]{g97} Galvin, A. B. 1997, Geophysical Monograph 99,
      ed. N. Crooker, J. Joselyn, \& J. Feynman, (AGU:Washington, D.C.), 253
\bibitem[Gloeckler et al.(1999)]{gla99} Gloeckler et al. 1999, 
      Geophys. Res. Lett., 26, 157
\bibitem[Kharchenko \& Dalgarno(2000)]{kd00} Kharchenko, V., \& Dalgarno, A., 
      2000, \jgr, 105(A8), 18,351
\bibitem[Kharchenko et al.(2003)]{kea03} Kharchenko, V., Rigazio, M., 
     Dalgarno, A., \& Krasnopolsky, V. A. 2003, \apj, 585, L73
\bibitem[Kuntz \& Snowden(2000)]{ks00} Kuntz, K., \& Snowden, S. L. 2000,
     \apj, 543, 195
\bibitem[Kuntz et al.(2004)]{kea04} Kuntz, K., et al. 2004, \apj, in preparation
\bibitem[Lallement(2004)]{lal04} Lallement, R. 2004, \aap, in press
\bibitem[Lisse et al.(1996)]{lis96} Lisse, C. M. et al. 1996, Science, 274, 205
\bibitem[Lisse et al.(2001)]{lea01} Lisse, C. M., Christian, D. J., 
      Dennerl, K., Meech, K. J., Petre, R., Weaver, H. A., and Wolk, S. J. 
      2001, Science, 292, 1343
\bibitem[McCammon et al.(2002)]{mea02} McCammon, D., et al. 2002, \apj, 576, 188
\bibitem[Petrinec \& Russell(1996)]{pr96} Petrinec, S. M., \& 
     Russell, C. T. 1996, \jgr, 101, 137
\bibitem[Richardson \& Paularena(2001)]{rp01} Richardson, J. D., \& 
     Paularena, K. I. 2001, \jgr, 106, 239
\bibitem[Robertson \& Cravens(2003a)]{rc03a} Robertson, I. P., \& Cravens, T. E. 
     2003a, Geophys. Res. Lett., 30(8), 1439
\bibitem[Robertson \& Cravens(2003b)]{rc03b} Robertson, I. P., \& Cravens, T. E. 
     2003b, \jgr, 108 (A10), 8031
\bibitem[Rucinski et al.(1996)]{rea96} Rucinski, D., Cummings, A. C., 
Gloeckler, G., Lazarus, A. J., Mobius, E., \& Witte, M. 1996, Space Sci. Rev., 
     78, 73
\bibitem[Schwadron \& Cravens(2000)]{sc00} Schwadron, N. A., \& Cravens, T. E. 
     2000, \apj, 544, 558
\bibitem[Smith et al.(2003)]{smea03} Smith, E. J. et al. 2003, Science, 302, 1165
\bibitem[Snowden et al.(1994)]{sea94} Snowden, S. L., McCammon, D.,
     Burrows, D. N., \& Mendenhall, J. A. 1994, \apj, 424, 714
\bibitem[Snowden et al.(1995)]{sea95} Snowden, S. L., Freyberg, M. J.,
     Schmitt, J. H. M. M., Voges, W., Tr\"umper, J. , Edgar, R. J.,
     McCammon, D., Plucinsky, P. P., \& Sanders, W. T. 1995,
     \apj, 454, 643
\bibitem[Snowden et al.(1997)]{sea97} Snowden, S. L., Egger, R., 
     Freyberg, M. J., McCammon, D., Plucinsky, P. P., Sanders, W. T., 
     Schmitt, J. H. M. M., Tr\"umper, J., \& Voges, W. 1997, \apj, 485, 125
\bibitem[Snowden et al.(1998)]{sea98} Snowden, S. L., Egger, R., 
     Finkbeiner, D. P., Freyberg, M. J., \& Plucinsky, P. P. 1998, 
     \apj, 493, 715
\bibitem[Wargelin et al.(2004)]{wea04} Wargelin, B. J., Markevitch, M., Juda, M., 
      Kharchenko, V., Edgar, R. J., \& Dalgarno, A. 2004, \apj, submitted
\end{thebibliography}
\end{document}